\documentclass[prb,twocolumn,showpacs,superscriptaddress,amsmath,amssymb]{revtex4}
\usepackage{graphicx}

\usepackage{dcolumn}
\usepackage{bm}
\usepackage{color}

\begin{document}

\title{Photoinduced coherent oscillations in the one-dimensional two-orbital Hubbard model}

\author{N. Maeshima}
\email{maeshima@ims.tsukuba.ac.jp}
\affiliation{Institute of Materials Science, University of Tsukuba, Tsukuba 305-8573, Japan}
\affiliation{Center for Computational Sciences, University of Tsukuba, Tsukuba 305-8577, Japan}

\author{K. Hino}
\affiliation{Institute of Materials Science, University of Tsukuba, Tsukuba 305-8573, Japan}
\affiliation{Center for Computational Sciences, University of Tsukuba, Tsukuba 305-8577, Japan}

\author{K. Yonemitsu}
\affiliation{Institute for Molecular Science, Okazaki 444-8585, Japan}
\affiliation{Department of Functional Molecular Science, Graduate University for Advanced Studies, Okazaki 444-8585, Japan}
\affiliation{CREST, JST, CREST, Tokyo 102-0075, Japan}

\date{\today}

\begin{abstract}
We study photoinduced ultrafast coherent oscillations originating from orbital degrees of freedom in the one-dimensional two-orbital Hubbard model.
By solving the time-dependent Schr\"{o}dinger equation for the numerically exact many-electron wave function,
we obtain time-dependent optical response functions.
The calculated spectra show characteristic coherent oscillations that vary with the frequency of probe light.
A simple analysis for the dominant oscillating components clarifies that 
these photoinduced oscillations are caused by the quantum interference between photogenerated states.
The oscillation attributed to the Raman-active orbital excitations (orbitons) clearly appears around the charge-transfer peak.
\end{abstract}

\maketitle


Photoinduced phenomena of strongly correlated electron systems have attracted much attention recently.~\cite{nasu,tokura,yonemitsu1}
For example, there have been many studies on photoinduced macroscopic changes in electronic states, often called ``photoinduced phase transitions'' (PIPTs).~\cite{TTFCA_1,poly,MX_1,PrMnO3_1,VO2_1,GdSrMnO3}

These photoinduced phenomena often accompany subsequent non-equilibrium dynamics.  One typical example is coherent oscillations observed after the rapid photoinduced changes.~\cite{TTFCA_2,EDO,KTCNQ2,MX_2,VO2_2,PrMnO3_3,LaMnO3,NdCaMnO3}
These oscillations involve much information of characteristic collective modes of the systems, phonon, orbiton, and so on.  Hence investigating the coherent oscillations provides us insight into roles of these modes in the photoinduced phenomena.

Until a few years ago, experimental studies have used relatively long pulses ($\sim$ 100fs), which allow us to detect only slow lattice dynamics.~\cite{TTFCA_2,EDO,KTCNQ2,MX_2}
However, recent development of experimental technique that provides sub-10-fs pulses enables us to observe much faster dynamics.
In particular, considerable experimental effort has been devoted to the study of the ultrafast oscillations in transition metal oxides,
which have fast vibrational phonon modes,~\cite{VO2_2,PrMnO3_3,LaMnO3,NdCaMnO3} or orbital excitations.~\cite{PrMnO3_3}

In contrast to these experimental achievements, 
theoretical studies on the photoinduced ultrafast oscillations have not been carried out so intensively.~\cite{yonemitsu2,onda,lee}
Although some of the authors and coworkers have provided a theoretical description on dynamics of an organic compound (EDO-TTF)$_2$PF$_6$, the treatment for the lattice degrees of freedom is limited to a classical one.~\cite{yonemitsu2,onda}
A quantum theory for the same material,~\cite{lee} where quantized phonons are dealt with, focuses on the slow lattice dynamics.  Thus alternative quantum-mechanical treatment is needed to describe the ultrafast oscillations of excitations with much higher frequencies.

\begin{figure}[ht]
\includegraphics[width=5.9cm,clip]{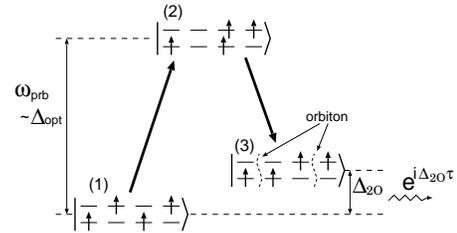}
\caption{The excitation process contributing to the coherent oscillation in the CT region.
The bold solid arrows show the optical transitions connecting the initial antiferro-orbital state (1) and the final two-orbiton state (3) via the intermediate optically excited state (2).
The quantum interference between (1) and (3) generates the coherent oscillation with the frequency $\Delta_{\rm 2O}$,
the energy gap between the two states.
A clear oscillation appears in the case where the probe frequency $\omega_{\rm prb}$ is almost equal to the optical gap $\Delta_{\rm opt}$.}
\label{fig_orbiton}
\end{figure}

In this Rapid Communication, we present a theoretical study of the photoinduced ultrafast coherent oscillations of the one-dimensional (1D) two-orbital Hubbard model coupled with static lattice distortion, which is a 1D analog of transition-metal oxides with orbital degrees of freedom.
Numerically calculated time-dependent optical response functions show clear ultrafast coherent oscillations
 that vary with the frequency of probe light.
An analysis of optical excitation processes contributing the dominant oscillating components clarifies that
(i) the photoinduced coherent oscillations are caused by the quantum interference between eigenstates included in the photoexcited state and
(ii) the oscillation around the charge-transfer (CT) peak results from the Raman-active two-orbiton state while the oscillation in the low-energy region is caused by a one-holon-doublon (hd)-pair excitation.
The excitation process for the former case is illustrated in Fig.~\ref{fig_orbiton}.

In this work, we use the 1D two-orbital Hubbard model coupled with static lattice distortion.
The Hamiltonian is given by
\begin{eqnarray}
{\cal H} &=& -\sum_{l\tau\sigma}[ t(\tau)  c^\dagger_{l\tau\sigma}c_{l+1\tau\sigma} +  {\rm H.c.}] + U\sum_{l\tau} n_{l\tau\uparrow}n_{l\tau\downarrow}  \nonumber \\
&+&U'\sum_{l} n_{l1}n_{l2} + J\sum_{l\sigma\sigma'} c^\dagger_{l1\sigma}c^\dagger_{l2\sigma'}c_{l1\sigma'}c_{l2\sigma} \nonumber \\ &+& J'\sum_{l,\tau\ne \tau'} c^\dagger_{l\tau\uparrow}c^\dagger_{l\tau\downarrow}c_{l\tau'\downarrow}c_{l\tau'\uparrow}  - g\sum_l Q_l (n_{l1}-n_{l2}) \nonumber\\
&+& \frac{K}{2} \sum_l Q_l^2, 
\label{eq_ham}
\end{eqnarray}
where $c^{\dagger}_{l\tau\sigma}$ ($c_{l\tau\sigma}$) is the creation (annihilation) operator of an electron with spin $\sigma$($=\uparrow,\downarrow$) at orbital $\tau$($=1,2$) at site $l$, $n_{l\tau\sigma}=c^{\dagger}_{l\tau\sigma}c_{l\tau\sigma}$, $n_{l\tau} = n_{l\tau\uparrow} + n_{l\tau\downarrow}$, and $Q_l$ is the Jahn-Teller-type lattice distortion.
$U$, $U'$, $J$, and $J'$ denote intraorbital Coulomb, interorbital Coulomb, interorbital spin exchange, and interorbital pair hopping interactions, respectively.
We also note that the following relations $U = U' + 2J$ and $J'=J$ hold.~\cite{kanamori}
The electron-lattice coupling and the elastic constant are given by $g$ and $K$, respectively.
We treat the quarter-filled $N$-site chain with $N=4$ and impose the periodic boundary condition.

The time($\tau$)-dependent transfer integral $t(\tau)$, which is finite only between the same orbitals of neighboring sites, is introduced as $t(\tau)= t_0 e^{i (ae/\hbar c) A(\tau)}$,
where $t_0$ is the bare transfer integral, $e$ is the absolute value of the electronic charge, $a$ is the lattice spacing, and $c$ is the velocity of light.  In the following, we use the unit $t_0=e=a=c=\hbar=1$.
The pump laser pulse is represented by the vector potential $A(\tau)$ given by
\begin{equation}
  A(\tau) = \frac{F}{\omega_{\rm pmp}} \cos(\omega_{\rm pmp}\tau) \frac{1}{\sqrt{2\pi}T_{\rm pmp}}e^{ - \frac{(\tau - \tau_{\rm c})^2}{2T_{\rm pmp}^2} },
\end{equation}
where $F$ is the amplitude of the electric field, $\tau_{\rm c}$ is the central time of the pump field, and $T_{\rm pmp}$ defines the width of the Gaussian function.
We set $\tau_{\rm c}=10$ and $T_{\rm pmp}=1$.
When we set the bare transfer integral $t_0=0.1$ eV, the pulse width $2T_{\rm pmp}$ corresponds to about 13 fs, which is same order of the pulse width of recent experiments.~\cite{VO2_2,PrMnO3_3,LaMnO3,NdCaMnO3}
The frequency $\omega_{\rm pmp}$ is set to the optical gap $\Delta_{\rm opt}$.

The procedure of calculation is as follows.   First of all, we obtain the ground state $|\phi_0\rangle$ and the stable lattice distortion $Q_l$ with no pump field by iterative application of the Lanczos diagonalization and the Hellmann-Feynmann theorem;~\cite{maeshima1}
$\partial\langle \phi_0 |{\cal H}| \phi_0\rangle/\partial Q_l = 0$ for all $l$.
The obtained stable configuration is found to be the staggered distortion $Q_l=(-1)^l Q_{\rm st}$ for $l=[0,\cdots,N-1]$.
Then, we calculate the state $|\psi(\tau)\rangle$ by solving the time-dependent Schr\"{o}dinger equation $i \frac{d}{d\tau}|\psi\rangle = {\cal H}|\psi\rangle$.
The ground state $|\phi_0\rangle$ is used as the initial state and the lattice configuration is fixed.
The Schr\"{o}dinger equation is numerically solved by expanding the exponential time evolution operator with time slice $d\tau$=0.02.~\cite{yonemitsu2}

To observe the time-dependent dynamics, we calculate the transient optical response function~\cite{matsueda,onda} given by
\begin{equation}
  {\rm I}(\omega_{\rm prb},\tau) = -\frac{1}{N} {\rm Im} \langle\psi(\tau)|\hat{j} \frac{1}{\omega_{\rm prb} + i\delta + E - {\cal H}_0}
  \hat{j}|\psi(\tau)\rangle,
  \label{eq_Iw}
\end{equation}
where $\hat{j}=it_0\sum_{l\tau\sigma}(c^\dagger_{l\tau\sigma}c_{l+1\tau\sigma} - c^\dagger_{l+1\tau\sigma}c_{l\tau\sigma} )$ is the current operator,~\cite{ss} $\delta$ is a broadening parameter set at 2.0, 
${\cal H}_0$ is Hamiltonian~(\ref{eq_ham}) with $t(\tau)=t_0$, and $E=\langle \psi(\tau)|{\cal H}_0|\psi(\tau)\rangle$.
We also note that $\Delta_{\rm opt}$ is obtained from the lowest peak of $I(\omega_{\rm prb},\tau)$ at $\tau=0$.

We set other parameters as $U'=20$, $J=5$, $g=0.4$, and $K=1$.  For these parameters the ground state is in the ferromagnetic phase with finite lattice distortion $Q_{\rm st}\sim 0.36$.
Because we focus on the photoinduced coherent oscillation phenomena in this work, the pump field is set weak, $F$=2.0, which does not cause a PIPT.


\begin{figure}[ht]
\includegraphics[width=5.3cm,clip]{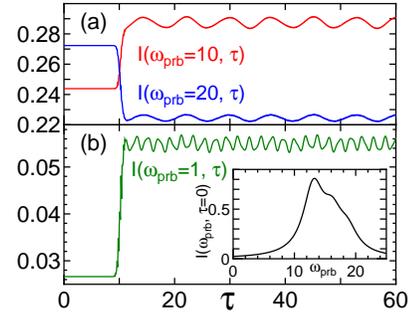}
\caption{(Color online). The optical response function ${\rm I}(\omega_{\rm prb},\tau)$ for (a) $\omega_{\rm prb}=$ 10
and 20, and (b) for $\omega_{\rm prb}=1$. The inset of (b) shows the optical response ${\rm I}(\omega_{\rm prb},\tau=0)$.}
\label{fig_swt}
\end{figure}

Figure~\ref{fig_swt} (a) shows the optical response ${\rm I}(\omega_{\rm prb},\tau)$ for $\omega_{\rm prb}=$ 10 and 20, which are the lower and higher sides of the CT peak [see the inset of Fig.~\ref{fig_swt} (b)].
After the drastic change induced by the pump field, ${\rm I}(\omega_{\rm prb}=10,20,\tau)$ shows clear coherent oscillations.
Their periods are about $8$ for both $\omega_{\rm prb}$=10 and 20, which suggests that their origins are the same low-lying excitations.
By contrast, the oscillation for ${\rm I}(\omega_{\rm prb}=1,\tau)$ with period $\sim 2$ [see Fig.~\ref{fig_swt} (b)] 
is evidently caused by other excitations. 

To clarify the origins of the coherent oscillations around the CT peak, we calculate the Fourier transform of $I(\omega_{\rm prb},\tau)$ and two spectral functions defined below.  The Fourier transform in the time domain $\tau\in[\tau_i,\tau_e]$ is given by
\begin{equation}
 \bar{I}(\omega_{\rm prb},\omega) = \left| \frac{1}{\sqrt{2\pi}} \int_{\tau_i}^{\tau_e} d\tau e^{i\omega\tau} {\rm I}(\omega_{\rm prb},\tau)\right|.
\end{equation}
One of the spectral functions is that detects Raman-active excitations, defined by
\begin{equation}
  \chi(\omega) \equiv -\frac{1}{\pi} {\rm Im} \langle jj| \frac{1}{\omega + i\delta` + \epsilon_0 - {\cal H}_0} |jj\rangle,
  \label{eq_chi}
\end{equation}
where $|jj\rangle = \hat{j}\hat{j}|\phi_0\rangle - |\phi_0\rangle\langle\phi_0|\hat{j}\hat{j}|\phi_0\rangle$. The other one is the orbital dynamical structure factor $T^z(q,\omega)$ given by
\begin{equation}
  T^z(q,\omega) = -\frac{1}{N} {\rm Im} \langle \phi_0| T^z_{-q}\frac{1}{\omega + i\delta` + \epsilon_0 - {\cal H}_0} T^z_q|\phi_0\rangle,
\end{equation}
where $T^z_q = \sum T^z_l e^{-iql}$ and $T^z_l=(n_{l1}-n_{l2})/2$.

\begin{figure}[ht]
\includegraphics[width=5.3cm,clip]{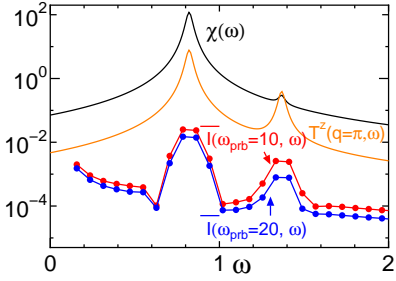}
\caption{(Color online). The Fourier transforms $\bar{I}(\omega_{\rm prb},\omega)$ for $\omega_{\rm prb}=$ 10 and 20 in the time domain $[\tau_i,\tau_e]=[20,100]$, and the spectral functions $\chi(\omega)$ and $T^z(q=\pi,\omega)$. The broadening parameter $\delta'$ is set at 0.05.}
\label{fig_fou}
\end{figure}

Figure~\ref{fig_fou} shows 
$\bar{I}(\omega_{\rm prb},\omega)$ for the time domain $[\tau_i,\tau_e]=[20,100]$, $\chi(\omega)$,
and $T^z(q=\pi,\omega)$.
All the functions have two distinct peaks: the dominant one with frequency $\omega = 0.8$, and the sub-dominant one with $\omega=1.4$.
Hence we conclude that the two peaks of $\bar{I}(\omega_{\rm probe},\omega)$ correspond to the Raman-active orbital excitations.
These orbital excitations are described by the effective model for the orbital degrees of freedom~\cite{kugel} given by
\begin{equation}
  {\cal H}_{\rm eff} = J \sum_l \vec{T}_l\cdot\vec{T}_{l+1} - H_{\rm eff} \sum_l (-1)^l T^z_l, 
\end{equation}
where $J$ is the antiferro-orbital superexchange constant, $J=4t_0^2/(U'-J)$, and $H_{\rm eff}$ is the effective staggered field, $H_{\rm eff}=2gQ_{\rm st}$.
The spin part is omitted since the system is ferromagnetic.  For the two-orbital system, the Raman-active orbital excitations have been discussed and shown that they are two-orbiton processes with a finite excitation gap due to the staggered field.~\cite{miyasaka}

\begin{figure}[ht]
\includegraphics[width=5.3cm,clip]{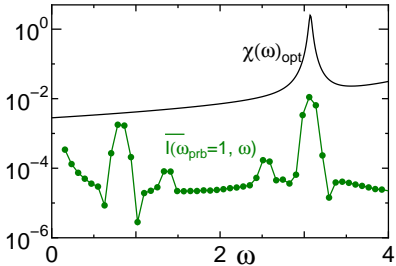}
\caption{(Color online). The Fourier transform $\bar{I}(\omega_{\rm prb},\omega)$ for $\omega_{\rm prb}=1$ in the time domain $[\tau_i,\tau_e]=[20,100]$ and $\chi(\omega)_{\rm opt}$ with $\delta'=0.05$.}
\label{fig_foulow}
\end{figure}

Now we turn our attention to the low-$\omega_{\rm prb}$ region.
 The Fourier transform for $\omega_{\rm prb}=1$, depicted in Fig.~\ref{fig_foulow}, displays the dominant peak at $\omega=3.1$, which is much higher than the frequencies of the orbital excitations.
The origin is elucidated by calculating an analog of $\chi(\omega)$ for the lowest optical
excitation $|\phi_{\rm opt}\rangle$ defined by
\begin{equation}
  \chi(\omega)_{\rm opt} \equiv -\frac{1}{\pi} {\rm Im} \langle jj'| \frac{1}{\omega + i\delta` + \epsilon_{\rm opt} - {\cal H}_0} |jj'\rangle,
  \label{eq_chi2}
\end{equation}
where $|jj'\rangle = \hat{j}\hat{j}|\phi_{\rm opt}\rangle - |\phi_{\rm opt}\rangle\langle\phi_{\rm opt}|\hat{j}\hat{j}|\phi_{\rm opt}\rangle$
and $\epsilon_{\rm opt}$ is the energy of $|\phi_{\rm opt}\rangle$.
The state $|\phi_{\rm opt}\rangle$ is the lowest one-holon-doublon(hd)-pair excitation,~\cite{maeshima2,maeshima3} which gives the main contribution of the low-$\omega_{\rm prb}$ component after photoexcitation.
$\chi(\omega)_{\rm opt}$ shown in Fig.~\ref{fig_foulow} has a clear peak at $\omega=3.1$, which is at the same location of that of $\bar{I}(\omega_{\rm prb}=1,\omega)$.
The state corresponding to this peak is another one-hd-pair excitation $|1{\rm hd}'\rangle$ in Fig.~\ref{fig_opt} (for details, see below), taking account of its eigenenergy; it is higher than that of $|\phi_{\rm opt}\rangle$ by the order of $t_0$ and much lower than the two-hd-pair excitations.

Now, let us discuss the reason why the different oscillations are observed by changing $\omega_{\rm prb}$.  To this end, we expand the quantum state $|\psi(\tau)\rangle$ as follows;
\begin{equation}
  |\psi(\tau)\rangle = \sum_{\alpha} C_{\alpha} e^{ -i \epsilon_{\alpha} \tau} |\alpha\rangle,
\end{equation}
where $|\alpha\rangle$ is an eigenstate of ${\cal H}_0$ and 
 has the eigenenergy $\epsilon_{\alpha}$.
We now assume that the total energy $E$ is almost equal to the ground-state energy $\epsilon_0$ since the pump field is weak in this work, and we thereby obtain the expression
\begin{eqnarray}
  I(\omega_{\rm prb},\tau) &=& \frac{1}{N}  \sum_{\alpha,\beta,\gamma} C^*_{\gamma} C_{\alpha} e^{ i(\epsilon_\gamma - \epsilon_\alpha)\tau } \nonumber \\ &\times&
  \langle \gamma|\hat{j} |\beta\rangle\langle \beta|\hat{j}|\alpha\rangle 
 f_{\rm L}(\omega_{\rm prb} + \epsilon_0 - \epsilon_\beta),
\label{eq_Iw2}
\end{eqnarray}
where $f_{\rm L}(x)$ is the Lorentzian function $f_{\rm L}(x) = \frac{\delta}{x^2 + \delta^2}$.
Equation~(\ref{eq_Iw2}) tells us the following points: 
(i) there are three important states, the initial state $|\alpha\rangle$, the final state $|\gamma\rangle$, and the virtually excited state $|\beta\rangle$, which are connected by the matrix element of $\hat{j}$.
 (ii) The coherent oscillation occurs as a quantum interference between $|\alpha\rangle$ and $|\gamma\rangle$ and its frequency is equal to the energy difference $\epsilon_\gamma-\epsilon_\alpha$.  (iii) The oscillation appears for $\omega_{\rm prb}\sim \epsilon_\beta - \epsilon_0$.

By using these points, we discuss the coherent oscillation around the CT gap, i.e.,  $\omega_{\rm prb} \sim \Delta_{\rm opt}=\epsilon_{\rm opt}-\epsilon_0$.
In this case, the relevant virtual state $|\beta\rangle$ is $|\phi_{\rm opt}\rangle$, and the important initial state is the ground state $|\phi_0\rangle$.
Then the final state is expected to be the two-orbiton state $|{\rm 2O}\rangle$, which is detected as the main peak of $T^z(q=\pi,\omega)$.
The schematic picture of this transition process is shown in Fig.~\ref{fig_opt} (a),
and the dominant component is given by
\begin{eqnarray}
  I(\omega_{\rm prb},\tau) &\sim& 
  \frac{1}{N} C^*_{{\rm 2O}} C_{0} e^{ i\Delta_{\rm 2O}\tau }   \langle {\rm 2O}|\hat{j} |\phi_{\rm opt}\rangle
\nonumber \\ 
&\times& \langle \phi_{\rm opt}|\hat{j}|\phi_{0}\rangle  f_{\rm L}(\omega_{\rm prb} -\Delta_{\rm opt}) + c.c. \ ,
\label{eq_Iw3}
\end{eqnarray}
where $\Delta_{\rm 2O}=\epsilon_{\rm 2O}-\epsilon_0$.
This expression clearly shows that the two-orbiton excitation $|2{\rm O}\rangle$ is observed as the main oscillating component and that $|2{\rm O}\rangle$ is Raman-active.

Now, let us discuss the low-$\omega_{\rm prb}$ region.  In this region the relevant initial state is the lowest optical excitation $|\phi_{\rm opt}\rangle$.
Then the main virtual state is the two-orbiton state $|{\rm 2O}\rangle$, and the final state should be another one-hd-pair state $|1{\rm hd}'\rangle$ [see Fig~\ref{fig_opt} (b)].
We thereby obtain the main contribution;
\begin{eqnarray}
  I(\omega_{\rm prb},\tau) &\sim& 
  \frac{1}{N}   C^*_{1{\rm hd'}} C_{\rm opt} e^{ i(\epsilon_{1{\rm hd}'} - \epsilon_{\rm opt})\tau }   \langle 1{\rm hd}'|\hat{j} |{\rm 2O}\rangle
\nonumber \\
&\times& \langle {\rm 2O}|\hat{j}|\phi_{\rm opt}\rangle f_{\rm L}(\omega_{\rm prb} -\Delta_{\rm 2O}) + c.c.\ .
\label{eq_Iw4}
\end{eqnarray}
From Eq.~(\ref{eq_Iw4}), we can see that the frequency shown in Fig.~\ref{fig_foulow} is equal to the gap between the lowest one-hd-pair state $|\phi_{\rm opt}\rangle$ and another one-hd-pair state $|1{\rm hd}'\rangle$.
In general, one-hd-pair states form a continuum in the thermodynamic limit,~\cite{Stephan_Penc} implying that an infinite number of coherent oscillations can contribute to $I(\omega_{\rm prb},\tau)$.
As a result, the coherent oscillation in the low-$\omega_{\rm prb}$ region may disappear because of the superposition of the infinite oscillating components.

\begin{figure}[ht]
\includegraphics[width=5.3cm,clip]{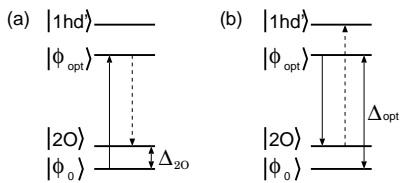}
\caption{The excitation processes contributing to the coherent oscillations (a) in the CT region and
(b) in the low-$\omega_{\rm prb}$ region.  The solid arrows show the optical transition processes from the initial state $|\alpha\rangle$ to the virtual state $|\beta\rangle$ and the dashed arrows show those from $|\beta\rangle$ to the final state $|\gamma\rangle$.}
\label{fig_opt}
\end{figure}

Here we discuss the experimental realization of the photoinduced oscillations caused by the orbital excitations.
The appearance of the photoinduced oscillation caused by the orbital excitations is suggested on the basis of the experimental results for Pr$_{0.7}$Ca$_{0.3}$MnO$_3$,~\cite{PrMnO3_3} a Mn perovskite with three-dimensional structure.  However, there is a puzzling fact that the oscillation is observed only above the orbital melting temperature.
Here, we note that our theory does not prohibit such an oscillation if Raman-active orbital excitations exist in the disordered phase.  In addition, if other excitations such as phonons have dominant Raman intensity, it might be difficult to distinguish the orbital excitations by using the Fourier transformation even in the ordered phase.

A more appropriate candidate of the quasi-1D system is LaVO$_3$,~\cite{miyasaka2,motome} where the Raman-active two-orbiton excitations exist.~\cite{miyasaka}  As for the photoinduced properties, no coherent oscillation has been detected while a photoinduced Drude-type spectral weight has been observed.~\cite{tomimoto}  Nonetheless, the femtosecond time-resolved reflection spectroscopy would clarify the oscillations because there found the Raman-active orbital excitations with the frequencies of 43 and 62 meV,~\cite{miyasaka} which correspond to oscillations with the time-periods of 96 and 67 fs.  Other quasi-1D materials, including KCuF$_3$,~\cite{kadota} would be alternative candidates with orbital degrees of freedom.


In summary, our quantum-mechanical treatment provides a simple picture for the photoinduced ultrafast coherent oscillations; the oscillations observed in the optical response are caused by the quantum interference between the eigenstates included in the photoexcited state.  The difference of the virtual optical excitation process results in the $\omega_{\rm prb}$ dependence of the oscillations.


This work was supported by Grants-in-Aid for Scientific Research on Innovative Areas ``Optical science of dynamically correlated electrons (DYCE)'' (Grant No. 21104504) and for Scientific Research (C) (Grant No. 19540381) and by ``Grand Challenges in Next-Generation Integrated Nanoscience'' from MEXT, Japan. 
Numerical calculations were carried out on T2K-Tsukuba System in Center for Computational Sciences in University of Tsukuba, and PrimeQuest in Research Center for Computational Science, Okazaki, Japan.




\end{document}